\documentclass[useAMS,usenatbib]{mn2e}
\usepackage{graphicx}
\usepackage{amsmath}
\usepackage{color}
\def\msun{M$_{\odot}$}
\def\Msun{M$_{\odot}$ }
\def\be{\begin{equation}}
\def\ee{\end{equation}}
\def\bi{\begin{itemize}}

\def\ei{\end{itemize}}
\def\ben{\begin{enumerate}}
\def\een{\end{enumerate}}
\def\bea{\begin{eqnarray}}
\def\eea{\end{eqnarray}}
\def\bt{\begin{tabbing}}
\def\et{\end{tabbing}}
\def\gcc{gcm$^{-3}$}
\def\edo{\end{document}}

\def\ye{$Y_e$}
\def\Ye{$Y_e$ }

\def\M4{M$_4$}
\def\M6{M$_6$}
\def\lcm6{LCM$_6$}
\def\qcm6{QCM$_6$}

\title[Robust r-process in neutron star mergers] {On the astrophysical
  robustness of neutron star merger r-process} \author[O. Korobkin, et
al.]{O. Korobkin$^{1}$\thanks{E-mail:
    o.korobkin@jacobs-university.de},
  S. Rosswog$^{1,2,3}$\thanks{E-mail: s.rosswog@jacobs-university.de},
  A. Arcones$^{4,5}$\thanks{E-mail: almudena.arcones@physik.tu-darmstadt.de}
  C. Winteler$^{6}$\thanks{E-mail: christian.winteler@unibas.ch}\\
  $^{1}$School of Engineering and Science, Jacobs University Bremen, Germany\\
  $^2$TASC, Department of Astronomy and Astrophysics, University of California, Santa Cruz, CA 95064,  USA\\
  $^3$Astronomy and Oskar Klein Centre, Stockholm University, AlbaNova, SE-10691 Stockholm, Sweden\\
  $^4$Institut f\"ur Kernphysik, Technische Universit\"at Darmstadt, Schlossgartenstra{\ss}e 2, D-64289 Darmstadt, Germany\\
  $^5$GSI Helmholtzzentrum f\"ur Schwerionenforschung GmbH, Planckstr. 1 D-64291 Darmstadt, Germany\\
  $^6$Department of Physics, University Basel, Switzerland}

\begin{document}

\date{Accepted 2012. Received 2012; in original form 2012}

\pagerange{\pageref{firstpage}--\pageref{lastpage}} \pubyear{2012}
\setlength\parindent{0pt}

\maketitle

\label{firstpage}

\begin{abstract}
In this study we explore the nucleosynthesis in the dynamic ejecta of compact binary 
mergers. We are particularly interested in the question how sensitive the resulting
abundance patterns are to the parameters of the merging system. Therefore, we systematically
investigate combinations of neutron star masses in the range from 1.0 to 2.0 \Msun and, 
for completeness, we compare the results with those from two simulations of a neutron star black hole 
merger. The ejecta masses vary by a factor of five for the studied systems, but
all amounts are (within the uncertainties of the merger rates) compatible with being 
a major source of cosmic r-process. The ejecta undergo a robust r-process nucleosynthesis 
which produces all the elements from the second to the third peak in close-to-solar 
ratios. Most strikingly, this r-process is extremely robust, all 23 investigated
binary systems yield practically identical abundance patterns. This is mainly the result
of the ejecta being extremely neutron rich (\ye $\approx0.04$) and the r-process 
path meandering along the neutron drip line so that the abundances are determined
entirely by nuclear rather than by astrophysical properties. While further questions related
to galactic chemical evolution need to be explored in future studies, we consider 
this robustness together with the ease with which both the second and third peak 
are reproduced as strong indications that compact binary mergers are prime 
candidates for the sources of the observed unique heavy r-process component.
\end{abstract}

\begin{keywords}
neutrinos -- nuclear reactions, nucleosynthesis, 
abundances 
\end{keywords}

\section{Introduction}
About half of the elements heavier than iron are formed by neutron capture reactions that
occur rapidly in comparison with $\beta$-decays. The basic physical 
mechanisms of this ``rapid neutron capture'' or ``r-process'', for short, had been 
identified already in the seminal paper of \cite{burbidge57}. Nevertheless, 
the cosmic cauldrons in which these heavy elements are forged have remained elusive for more than
half a century. Observations of metal-poor stars point to at least two groups of r-process events.
The first one is rare and produces whenever it occurs the heaviest r-process elements in nearly exactly
solar proportions. The second one, occurs more frequently and produces predominantly lighter 
elements from strontium to silver \citep{cowan06,honda06}. Given that its signature is 
less unique the second component might also be the result of a superposition from several 
sources (see e.g. \cite{arcones11c,frischknecht12}). Metal-poor stars that are enriched by r-process 
material show patterns that very closely match the (scaled) solar system abundances for 
nuclei beyond Ba (Z=56). This suggests that this heavy r-process component is produced only if 
a very unique set of astrophysical conditions is realized, or, alternatively, that it is produced 
in a range of realizations, but is insensitive to the exact parameters of its formation environment. 
So far, there is no generally accepted explanation for the unique heavy r-process component.\\
Historically supernovae were considered the ``natural'' source of r-process elements, but this
view has recently been distressed by a slew of investigations (e.g. 
\cite{arcones07,roberts10,fischer10,huedepohl10}).
These studies found neutrino-driven winds in supernovae to be seriously challenged in providing 
the physical conditions (high entropy, low electron fraction together with rapid expansion) 
that are required to produce the heavy ($A>90$) r-process elements. A possible exception may 
be magnetorotationally driven supernova jets where the fast ejection of highly compressed
neutron-rich matter produces interestingly low electron fraction values 
\citep{winteler12b}. It remains to be explored, however, how robust this 
scenario is with respect to the stellar parameters and with respect to its nucleosynthetic yields.
The main contenders of supernovae in terms of r-process nucleosynthesis are compact binary 
mergers of either two neutron stars (ns$^2$) or a neutron star and a stellar-mass black hole (nsbh)
\citep{lattimer74,lattimer76,eichler89,freiburghaus99b}.\\
%
  \begin{figure*}
  \begin{center}
  \centerline{
  \includegraphics[width=0.3\textwidth,angle=-90]{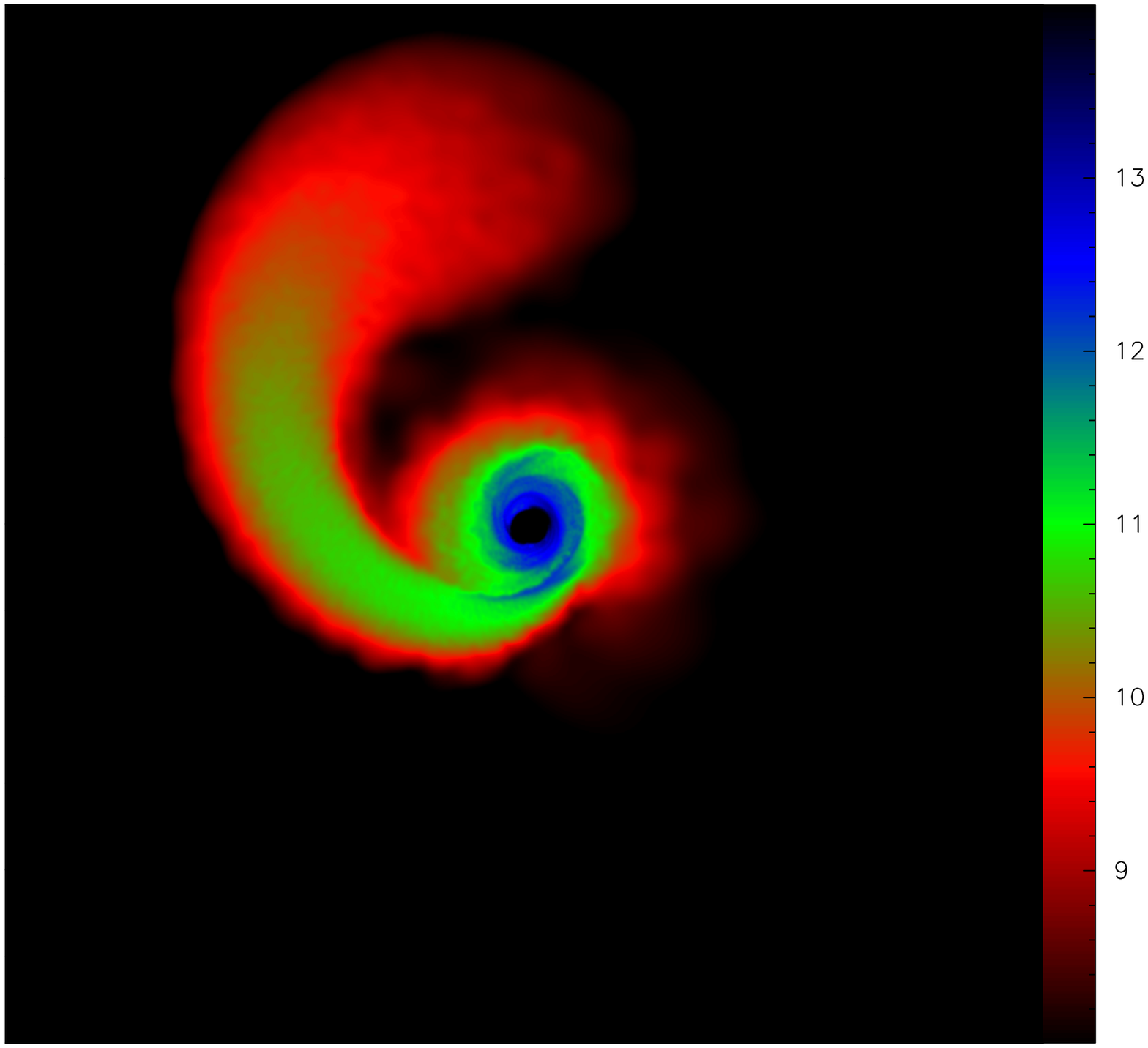}
  \includegraphics[width=0.3\textwidth,angle=-90]{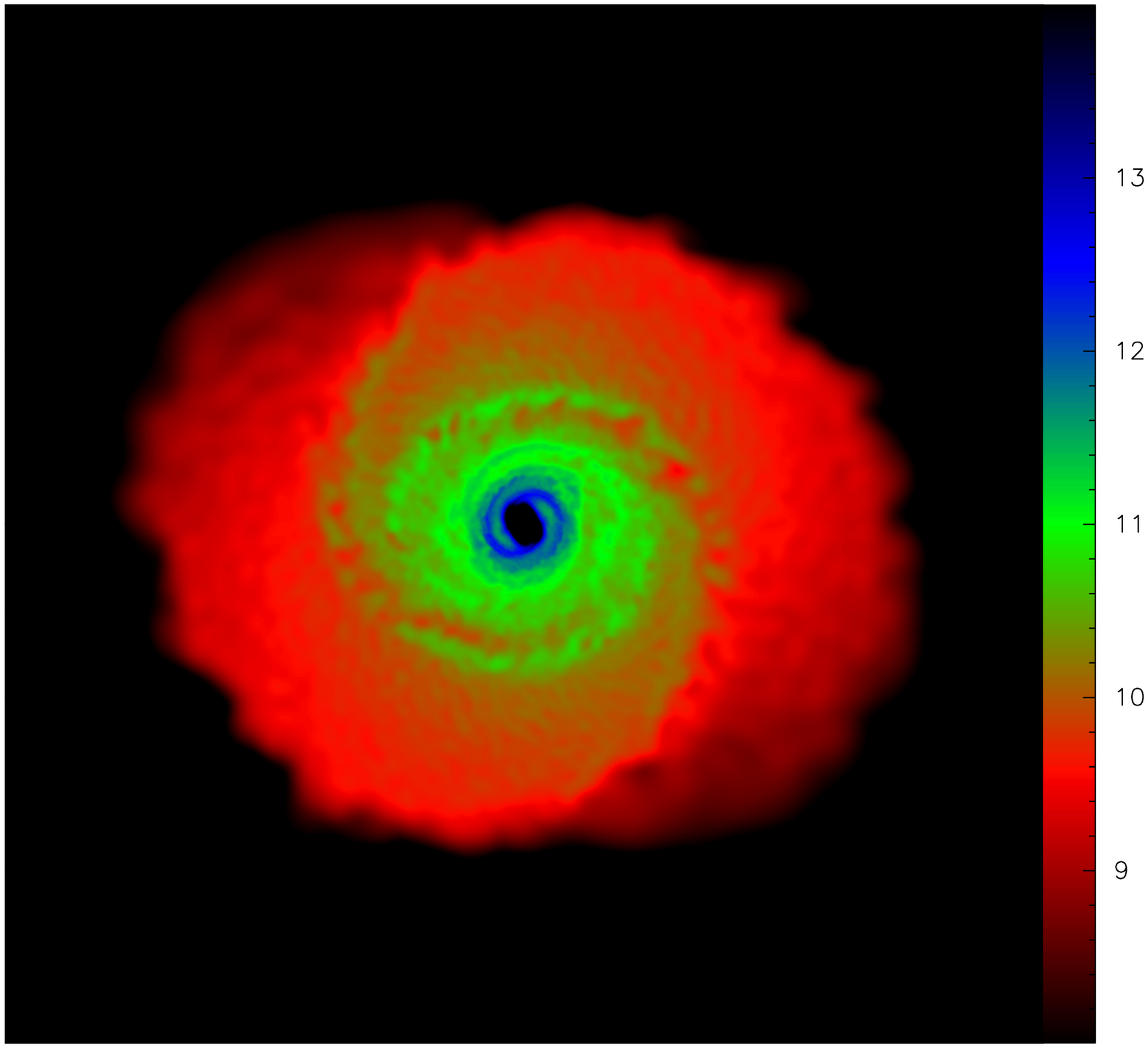}
  \includegraphics[width=0.3\textwidth,angle=-90]{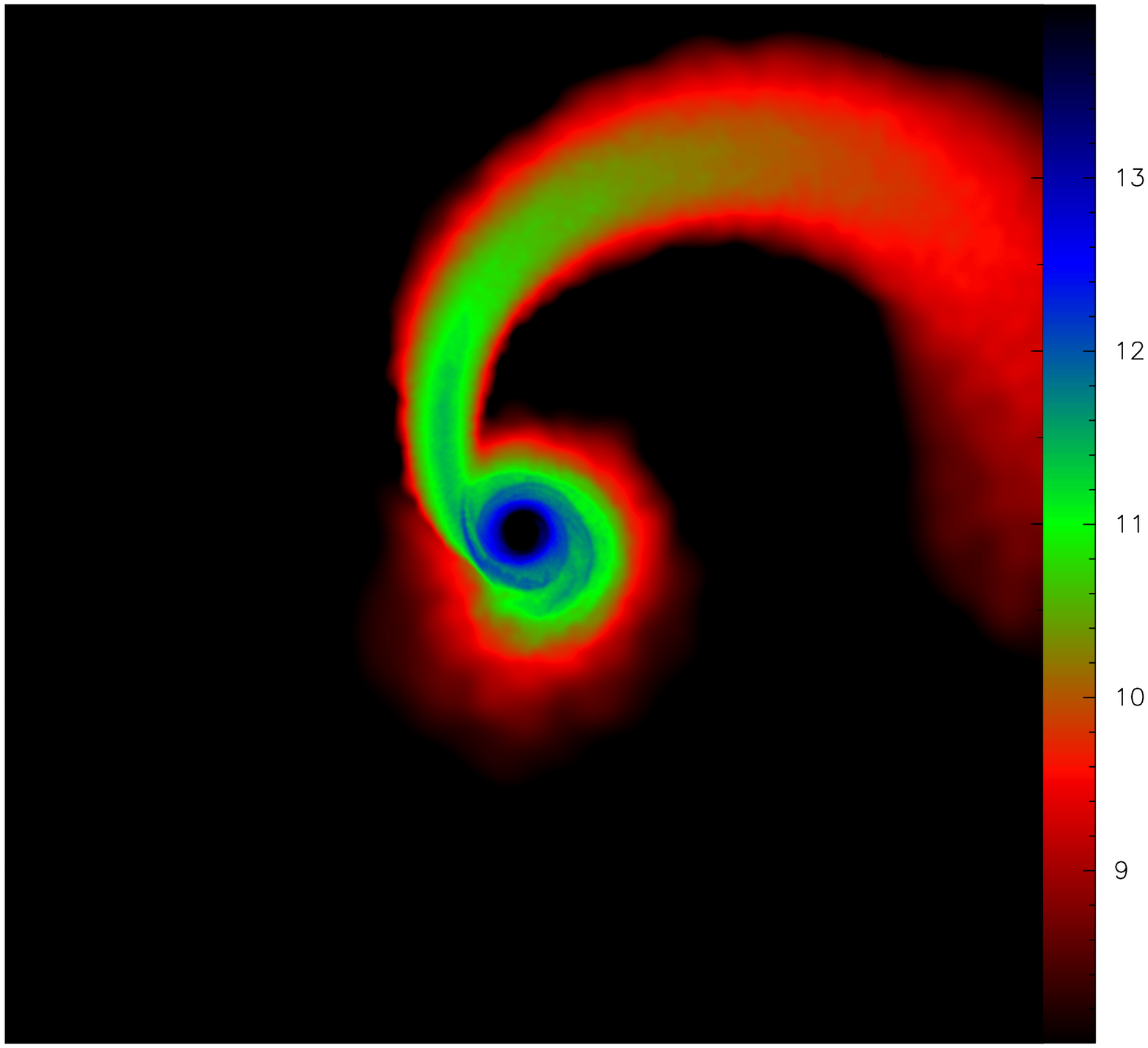} }
  \caption{Sensitivity to mass ratio: shown are density cuts ([600 km x 600 km], colour-coded
is the logarithm of density in cgs units) of 
    a 1.2-1.4 \Msun (t= 13.9 ms), a 1.4-1.4 \Msun (t= 13.4 ms) and a 2.0-1.4 \Msun (t= 15.0 ms)
    merger. 
  }
  \label{fig:mass_ratio_dependence}
  \end{center}
  \end{figure*}
Since the first nucleosynthesis calculations for neutron star merger ejecta \citep{freiburghaus99b} have been
performed, numerical models have seen vast improvements in physics, numerical methods and resolution and
we consider it timely to address this important topic again. There have been recent r-process studies in
a compact binary merger context \citep{roberts11,goriely11a,wanajo12} which, in agreement with the results 
of the first study, confirm the occurence of a robust r-process. While all previous studies provided 
new insights each of them had its limitations. For example, in all of them the ejecta electron fraction 
was poorly known since $Y_e$-changing weak interactions were ignored. More importantly, the ns$^2$ 
parameter space was explored only very punctually around ns masses of 1.4 \Msun(\cite{freiburghaus99b} 
explored one case from \cite{rosswog99}, two cases were explored in the conformal flatness approximation 
approach of \cite{goriely11a}\footnote{These authors also used the Lattimer-Swesty EOS to test for 
the impact of the neutron star equation of state.}
and three cases in \cite{roberts11}). In recent years, however, it has become clear that the neutron stars 
realized in nature span a substantial range in masses, see e.g. \cite{lattimer10} for  compilation of neutron star
masses. Therefore, an important open question needs to be answered: 
``How sensitive is the compact object merger nucleosynthesis to the actual astrophysical system that merges?''  \\
This is what we address in this paper: we systematically scan the neutron star binary parameter space in 21 
simulations and we calculate the resulting nucleosynthetic products. For completeness, we also show the 
results for the ejecta of two nsbh systems. Apart from exploring a very large neutron star mass range, 
we also significantly improve on the treatment of the electron fraction \Ye:
we start with cold neutron stars in $\beta$-equilibrium 
and allow for the evolution of $Y_e$ via electron and positron captures \citep{rosswog03a}.
We demonstrate that the nucleosynthesis results are essentially independent of the 
parameters of merging system, i.e. every ns$^2$ or nsbh system produces nearly exactly the same abundance pattern. 
The results show, however, some sensitivity to the properties  of extremely neutron-rich nuclei 
(e.g. binding energy, half-lives, fission properties) which are not well known \citep{thielemann11}.\\
Although heavy element nucleosynthesis is the clear focus of this study, we would like to stress the importance
of compact object merger ejecta and their nucleosynthesis in a broader astrophysical perspective. Both
the LIGO and Virgo gravitational wave (GW) detectors are currently being upgraded \citep{abbott09a,smith09,sengupta10} 
to a $\sim 10-15$ times better sensitivity than the original versions of the instruments. This will 
increase the volume of accessible astrophysical sources by more than a factor of 1000 reaching a detection 
horizon of a few hundred Mpc for ns$^2$ mergers and about a Gpc for nsbh mergers \citep{abadie10}. Since
the first detections will most likely be around or even below threshold it is of paramount importance
to identify accompanying electromagnetic signatures of the most promising GW sources to enhance the confidence
in a possible detection. Short gamma-ray bursts are likely the brightest electromagnetic manifestations that
result from a compact binary merger, but also the r-process nucleosynthesis and the decay of the resulting
radioactive nuclei may leave an observable signature as optical/UV transients, so-called ``macronovae''
\citep{li98,kulkarni05,rosswog05a,metzger10b,roberts11,metzger12a}. For neutron star mergers they are 
expected to peak
about 10 hours after the mergers with a few times $10^{42}$ erg/s \citep{rosswog12a,piran12a}. Later, on
time scales of months to years, the ejecta dissipate their kinetic energy in the ambient medium and thereby
produce possibly detectable radio flares \citep{nakar11a,piran12a}.\\
This paper is structured as follows. In Sec. \ref{sec:hydro} we briefly summarize our simulations and
explore the amount of ejected matter as a function of the binary system parameters. In Sec. \ref{sec:nucleo} 
we describe how we calculate the nuclear abundances and we discuss how they depend
on the astrophysical properties of the merging system and on the nuclear physics input.
In Sec. \ref{sec:discussion} we summarize our results and discuss their astrophysical implications.  

\section[]{Hydrodynamic evolution and dynamically ejected mass}
\label{sec:hydro}
\subsection{Explored parameter space}
Until recently, the neutron star mass distribution was thought to be clustered
narrowly around 1.35 \Msun \citep{thorsett99}. Therefore, essentially all neutron 
star merger studies have focused on a narrow range of masses around this value.
Recent observations, however, indicate a much broader neutron star mass
spectrum. There is now ample support for neutron star
masses significantly larger than 1.5 \msun. A broad peak around 1.5-1.7 \Msun
has been found \citep{kiziltan10,valentim11} for neutron stars with
white dwarf companions, an additional low-mass peak near 1.25 \Msun 
is thought to be characteristic for neutron stars that were produced by electron
capture supernovae \citep{podsiadlowski04,vandenheuvel04,schwab10}.
PSR J1614-2230 with 1.97 $\pm$ 0.04 \Msun possesses the largest accurately 
known neutron star mass  \citep{demorest10}, but even higher masses may exist in nature.
For example, the mass of the black widow pulsar has recently been determined as  2.4 
$\pm$ 0.12 \msun \citep{vankerkwijk11}. On the lower mass side, the secondary neutron star in J1518+4904 has a 
best estimated mass value of only 0.72 \msun, although with a very large 
1$\sigma$ error bar of 0.5 \msun, see \cite{lattimer10} and references therein.\\
We take these results as a motivation for a systematic exploration of 
the parameter space from  1.0 to 2.0 \Msun in steps of 0.2 \msun. 
Since the neutron star viscosity cannot substantially spin up the neutron stars
during the short tidal interaction phase preceeding the merger \citep{bildsten92,kochanek92}
all our models have vanishing initial ns spins. Our ns$^2$ cases are complemented 
by two nsbh cases with black hole masses of 5 and 10 \msun, for an overview
over the performed simulations see Tab.~\ref{tab:runs}. 
\begin{table*}
 \centering
 \begin{minipage}{140mm}
  \caption{Overview over the performed simulations, the superscript $^+$ indicates that the primary is a black hole.}
  \begin{tabular}{@{}rccccrl@{}}
  \hline
   Run   &  $m_1$ [\msun] & $m_2$ [\msun] & $N_{\rm SPH}\; [10^6]$ & $t_{\rm end}$ [ms] & $m_{\rm ej}$ [\msun]& comment \\
\hline 
\\
   1     &   1.0          & 1.0        & 1.0            &  15.3   & $ 7.64 \times 10^{-3}$ & \\
   2     &   1.2          & 1.0        & 1.0            &  15.3   & $ 2.50 \times 10^{-2}$ & \\
   3     &   1.4          & 1.0        & 1.0            &  16.5   & $ 2.91\times 10^{-2}$  & \\
   4     &   1.6          & 1.0        & 1.0            &   31.3  & $ 3.06\times 10^{-2}$  & \\
   5     &   1.8          & 1.0        & 1.0            &   30.4  & $>1.64 \times 10^{-2}$ & secondary still orbiting\\
   6     &   2.0          & 1.0        & 0.6            &   18.8  & $>2.39\times 10^{-2}$  & secondary still orbiting\\
\\
   7    &   1.2          & 1.2         & 1.0            &  15.4  & $ 1.68 \times 10^{-2}$ &  \\
   8    &   1.4          & 1.2         & 1.0            &  13.9  & $ 2.12 \times 10^{-2}$ &  \\
   9    &   1.6          & 1.2         & 1.0            &  14.8  & $ 3.33 \times 10^{-2}$ &  \\
   10   &   1.8          & 1.2         & 1.0            &  21.4  & $ 3.44 \times 10^{-2}$ &  \\
   11   &   2.0          & 1.2         & 0.6            &  15.1  & $>2.95 \times 10^{-2}$ &  secondary still orbiting\\   
\\
  12    &   1.4         & 1.4         & 1.0            &  13.4  & $1.28 \times 10^{-2}$  & \\
   13   &   1.6         & 1.4         & 1.0            &  12.2  & $2.36\times 10^{-2}$   & \\
   14   &   1.8         & 1.4         & 1.0            &  13.1  & $3.84\times 10^{-2}$   & \\
   15   &   2.0         & 1.4         & 0.6            &  15.0  & $3.89\times 10^{-2}$   & \\   
\\
   16   &   1.6         & 1.6          & 1.0            &  13.2  & $1.97 \times 10^{-2}$ & \\
   17   &   1.8         & 1.6          & 1.0            &  13.0  & $1.67 \times 10^{-2}$ & \\
   18   &   2.0         & 1.6          & 0.6            &  12.4  & $3.79\times 10^{-2}$  & \\
  \\
  19   &   1.8          & 1.8          & 1.0            &  14.0  & $1.50 \times 10^{-2}$ & \\
  20   &   2.0          & 1.8          & 0.6            &  11.0  & $1.99 \times 10^{-2}$ & \\
  \\
  21   &   2.0          & 2.0          & 0.2            &  21.4  &  $1.15\times 10^{-2}$ & \\  
  \\
  22   &   5.0 $^+$   & 1.4          &   0.2             & 138.7  &   $2.38 \times 10^{-2}$ & nsbh \\
  23   &  10.0 $^+$   & 1.4          &   0.2             & 139.0  &   $4.93 \times 10^{-2}$ & nsbh \\
  \hline \hline
\label{tab:runs}
\end{tabular}
\end{minipage}
\end{table*}

\subsection{Methodology}
To follow the hydrodynamic evolution we use the Smooth Particle Hydrodynamics (SPH) 
method, see \cite{monaghan05} and \cite{rosswog09b} for recent reviews, which, due to
its completely Lagrangian nature, is ideally suited to follow the ejected material.
Our code is an updated version of the one that was used in earlier studies
\citep{rosswog02a,rosswog03a,rosswog03c,rosswog05a}. It uses the Shen
et al.  equation of state (EOS) \citep{shen98a,shen98b} and an
opacity-dependent multi-flavour neutrino leakage scheme
\citep{rosswog03a}. The latter allows in particular to follow the \ye-changes
due to electron and positron captures. Moreover, we use a modern, time-dependent 
artificial viscosity prescription, see \cite{rosswog00,rosswog08b} for details.
The presented calculations make use of Newtonian gravity which leads
to less compact neutron stars than General Relativity. This could have 
a moderate effect on the amount of ejecta mass and, since the $\beta$-equilibrium
$Y_e$ is determined by the density inside the initial neutron star, the electron
fraction. As we will show below, however, the heavy nucleosynthesis 
is only affected once $Y_e$ reaches $\sim 0.2$ and the ejecta masses are (within
the existing uncertainties) still consistent with compact binary mergers
being a major r-process source even if the amount of ejecta is reduced 
by a factor of a few. Therefore, we expect our major results to be robust. 
Nevertheless this point needs improvement in future work.
  \begin{figure*}
  \centerline{
  \includegraphics[width=0.77\textwidth,angle=-90]{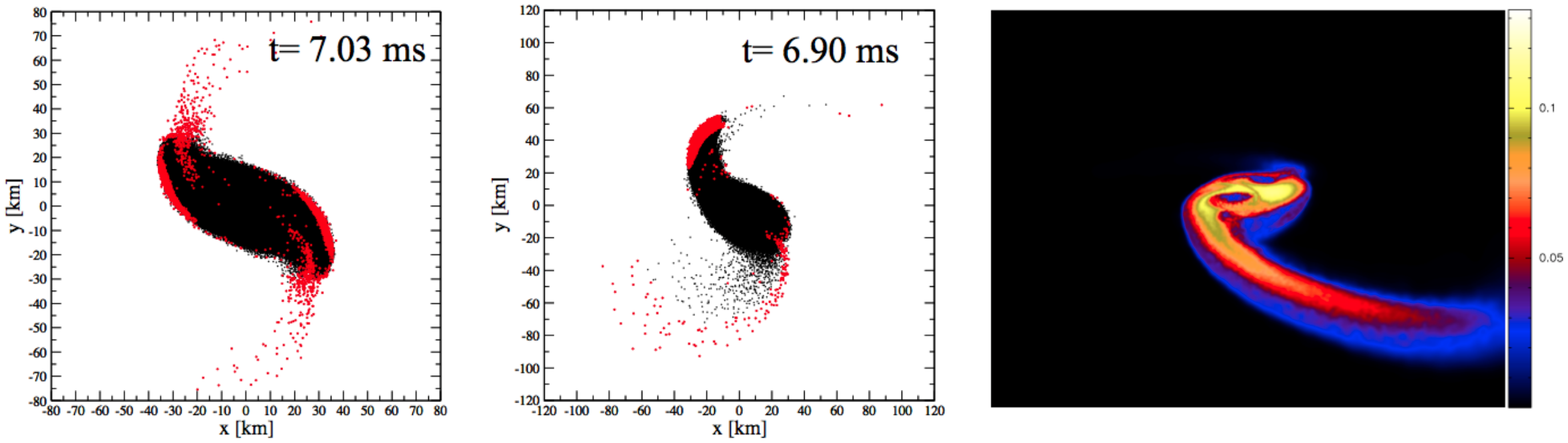}}
  \vspace*{-4cm}
  \caption{The ejecta come from two different regions (left and middle): a hot interaction region between
           the stars where matter is ejected by hydrodynamic effects  
           and a colder region that is flung out by tidal torques. The fraction of the 
           latter material increases with the asymmetry in the stellar masses (left 1.4 + 1.4 \msun,
           right 1.8 + 1.4 \msun). Right: volume rendering of the \Ye distribution (1.4 + 1.2 \Msun at
           t= 8.09 ms), only matter below the orbital plane is shown.}
  \label{fig:escaper_region}
  \end{figure*}
\subsection{Ejecta properties}
\begin{figure*}
  \centerline{
  \includegraphics[width=0.48\textwidth]{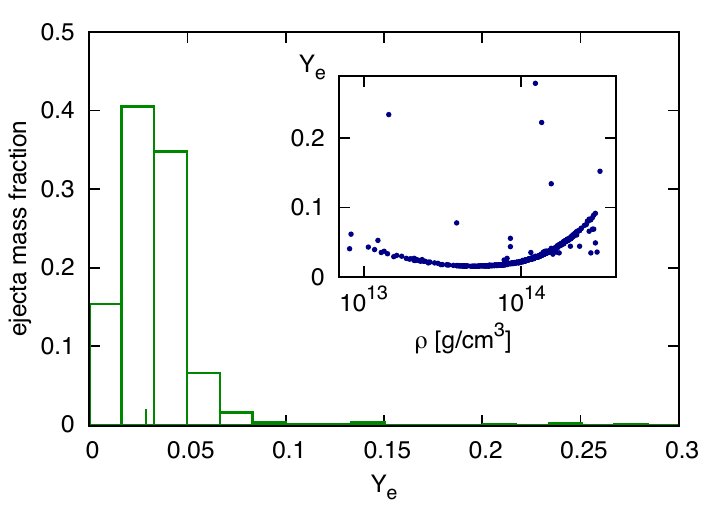}
  \includegraphics[width=0.5\textwidth,angle=0]{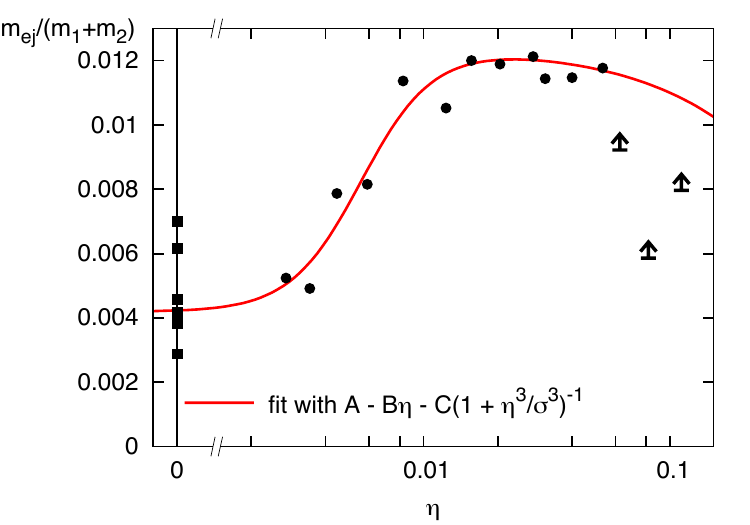} }
  \caption{Left panel: Distribution of \Ye in the selected subset of ejected particles in
  all ns$^2$ and nsbh simulations, binned by mass. 
  The inset shows how \Ye is correlated with the initial density in each
  particle before the merger.
  Majority of points on the $Y_e$-$\rho$ diagram trace the neutron star
  composition in the layer below the crust, where the $Y_e(\rho)$ reaches
  minimum.
  As a consequence, \Ye in the ejecta is narrowly distributed around
  $Y_e\sim0.04$.
  Right panel: Fit of the ejected mass normalized by the total mass of the binary
    as a function of the asymmetry parameter $\eta$ (see
    main text for definition).
    Coefficients of the fit are: $A=0.0125$, $B=0.015$, $C=0.0083$, and
    $\sigma=0.0056$.
    The arrows indicate the lower limits on the ejected mass for the three
    simulations in which the secondary is still not fully disrupted.
  }
  \label{fig:ye-hist}
  \end{figure*}

The merger dynamics is rather sensitive to asymmetries in the component masses,
see Fig.~\ref{fig:mass_ratio_dependence} (for a more complete overview 
over the remnant structures see
Fig. 1 in \cite{piran12a}). The asymmetries enhance the degree to which the 
secondary is disrupted and they lead to larger ejecta masses and velocities.
As already pointed out by \cite{oechslin07a}, the ejecta stem from two different regions: 
a hot component from the interaction region between the stars (subsequently called
``interaction-'' or i-component for short) and a colder one where matter is simply 
flung out by tidal torques, see the first two panels in 
Fig.~\ref{fig:escaper_region}. We refer to this matter as ``tidal'' or t-component. 
As the mass ratio decreases, the primary only ejects small amounts of the i-component, 
while the fraction and overall amount in the t-component from the secondary increases. 
This differs from the findings of \cite{goriely11a}, where even in unequal-mass cases 
most of the mass is lost from the primary as i-component. For now we can only speculate
about the origin of this difference: it could either be due the different treatment of
self-gravity (conformal flatness vs Newtonian) or, since the i-component comes from
a region of very strong shear, it could also to some extent be impacted by the 
specifics of numerical method, such as implementation of articifical viscosity.\\
As can be seen from the volume rendering in panel 3 of Fig.~\ref{fig:escaper_region},
the ejecta are extremely neutron-rich. In Fig.~\ref{fig:ye-hist}, left panel, we bin the mass fractions
of the ejecta of all simulations according to \ye. The values are clustered around \ye$= 0.03$
and dominated by t-component material. Since at high densities the \ye-changing reactions 
are Pauli-blocked and the subsequent rapid expansion leads to a very fast temperature drop,
the t-component had essentially no time to change its electron fraction. Therefore, it
still carries the original \ye-value set by cold, high-density $\beta$-equilibrium inside the 
neutron star. The \Ye of this matter therefore is a unique function of density which shows
up as well-defined curve in the inset of Fig.~\ref{fig:ye-hist}. The scattered component in 
this inset is a combination of matter from the i-region and matter from the high \Ye neutron 
star crust which is only marginally resolved in our simulations. The t-component of the ejecta 
not only starts within a narrow range of thermodynamical parameters, $Y_e=0.04\pm0.02$, 
$\rho = (1.4\pm0.5)\times10^{14}$~\gcc, corresponding to the outer $\sim 2$ km of the
original neutron star (see, for example, Fig. 1 in \cite{rosswog12a}), but also 
continues to expand adiabatically and without shocks.  Therefore, all fluid elements 
in tidal ejecta share approximately the same thermodynamic history and provide robust 
conditions for a uniform nucleosynthesis.\\
The amounts of ejected matter range from $7.6 \times 10^{-3}$ to $3.9
\times 10^{-2}$ \Msun ($4.9 \times 10^{-2}$ \Msun in the nsbh
case). We see a clear trend for decreasing mass ratios to eject more
matter, but for the equal-mass systems no unique trend is found with
respect to the total binary mass.  
For at least small ns mass asymmetries, i.e. for the most common cases,
we provide a fit formula based on the dimensionless mass 
asymmetry parameter 
\be
\eta:=1 -4m_1m_2/(m_1+m_2)^2.
\label{eq:eta}
\ee
Contrary to the mass ratio
$q$, $\eta$ is symmetric with respect to both masses and 
varies only in a finite range from $0$ to $1$ (with $0$ for a
equal-mass system and $\eta$ approaching 1 for extreme mass ratios)\footnote{
  A similar parameter $\bar{\eta}=m_1m_2/(m_1+m_2)^2$ was used in
  deriving an empirical formula for recoil velocities of coalescing
  non-spinning black holes~\citep{gonzalez07,fitchett83}
}.
The fit formula
\begin{align}
  m_{ej}(m_1,m_2) =
  (m_1+m_2)\left(A-B\eta-\frac{C}{1+\eta^3/\sigma^3}\right),
\end{align}
with $A=0.0125$, $B=0.015$, $C=0.0083$ and $\sigma=0.0056$, shown
as solid line in Fig.~\ref{fig:ye-hist}, right panel,
provides a reasonably good approximation to the simulation results.
For unequal-mass binaries $m_{ej}/(m_1+m_2)$ tends
to a minimum when approaching $\eta\to0$ and  a maximum of
$\approx0.12$ is reached near $\eta=0.02$.\\
\section{Nucleosynthesis}
\label{sec:nucleo}             
\setlength\parindent{0pt}
  \begin{figure}
  \begin{center}
  \includegraphics[width=0.48\textwidth]{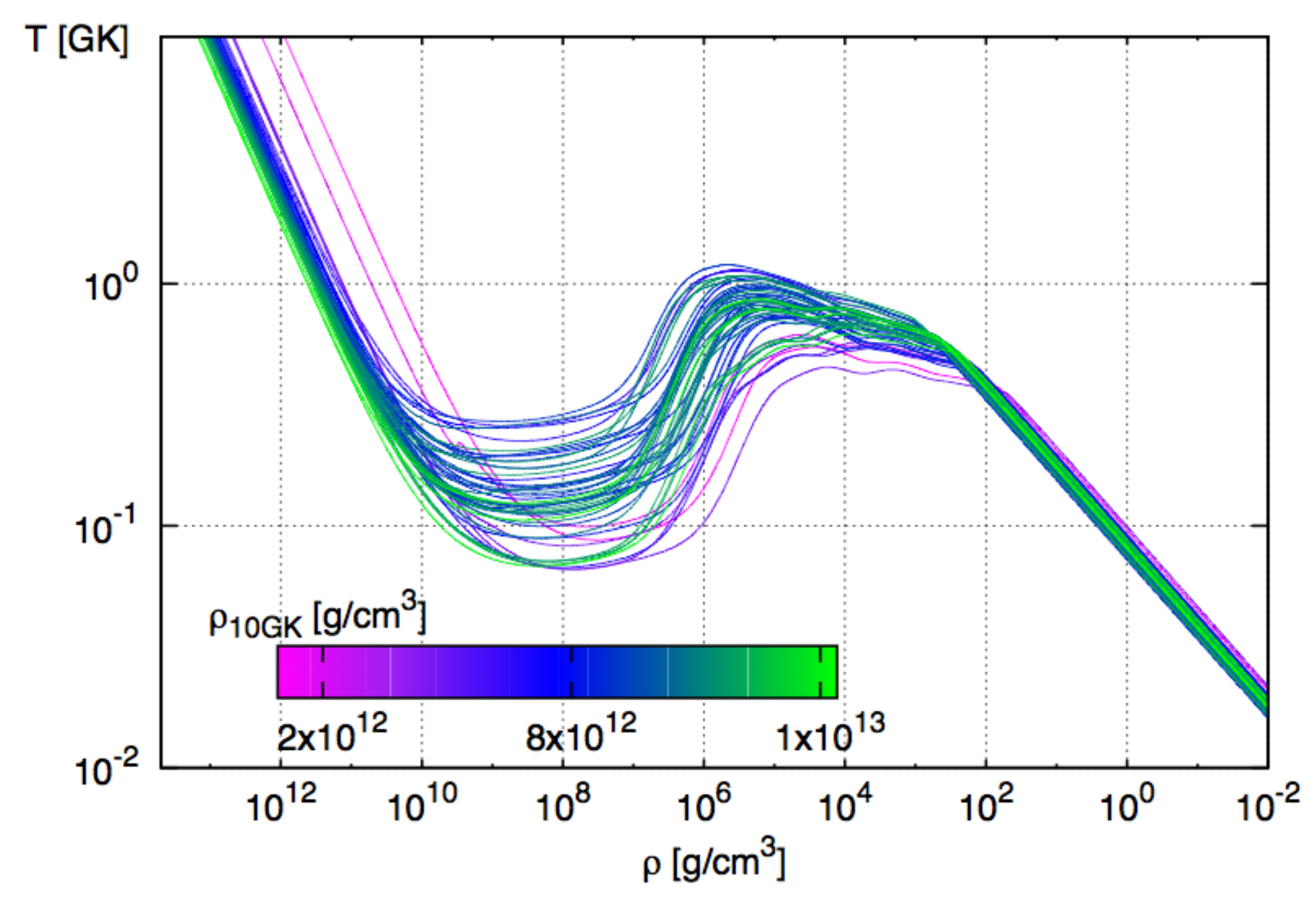}\\
  \includegraphics[width=0.48\textwidth]{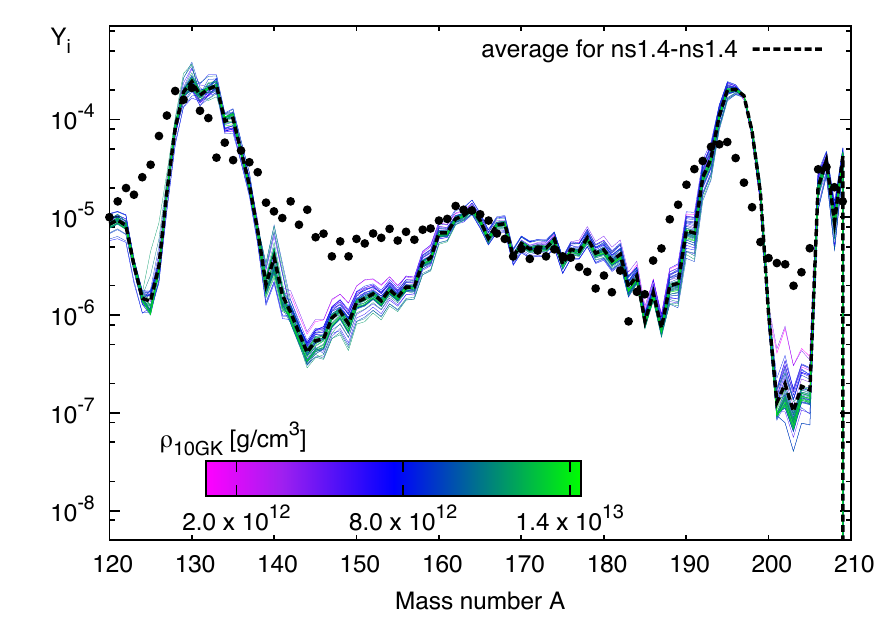}\\
  \includegraphics[width=0.48\textwidth]{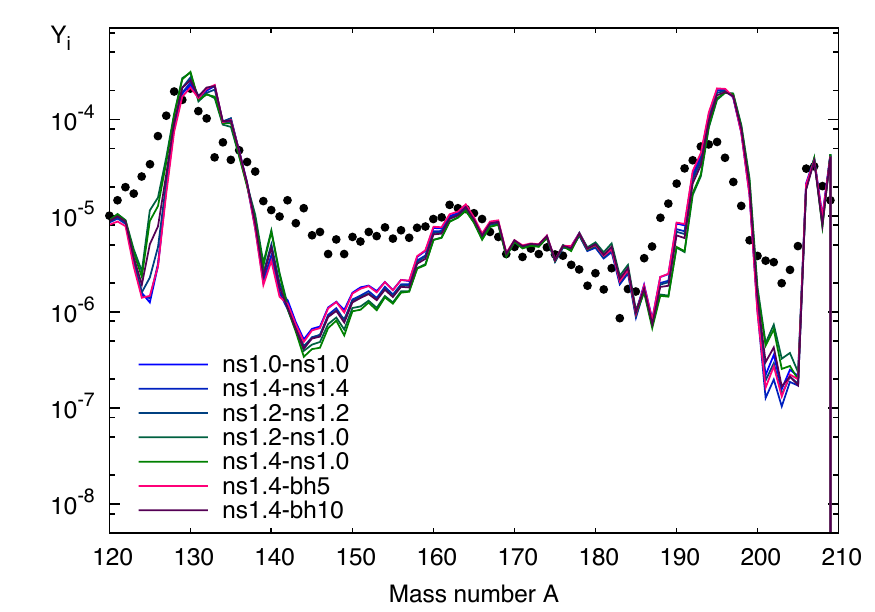}
  \caption{
    Top: density and temperature evolution for a bundle of trajectories,
    color-coded by density at $T=$ 10~GK.
    Middle: resulting final abundances distribution.
    Their averaged distribution is shown in a black dashed line, and a bold
    red line represents abundances for a trajectory without heating.
    All trajectories represent a subset from the standard 1.4-1.4~\Msun
    merger.
    Bottom:
    distribution of abundances for a variety of different (ns$^2$ and nsbh)
    merger cases. All different astrophysical systems yield essentially
    identical resulting abundances.
  }
  \label{fig:standard_results}
  \end{center}
  \end{figure}
\begin{figure}
  \begin{center}
    \includegraphics[width=0.48\textwidth]{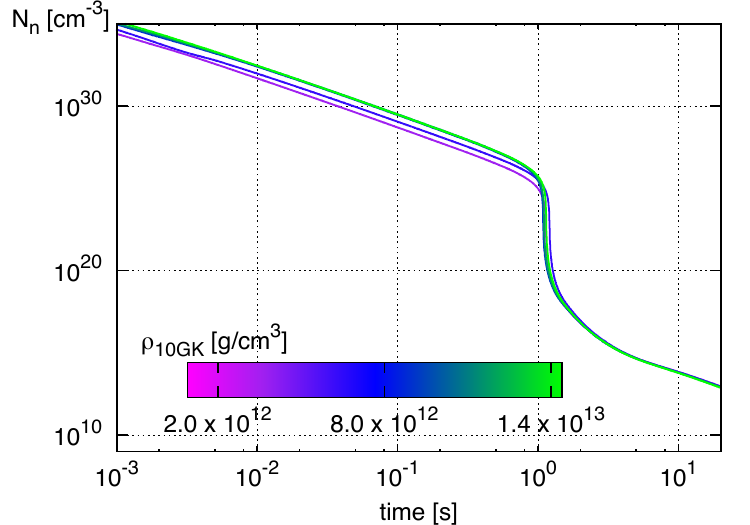}\\ 
    \includegraphics[width=0.48\textwidth]{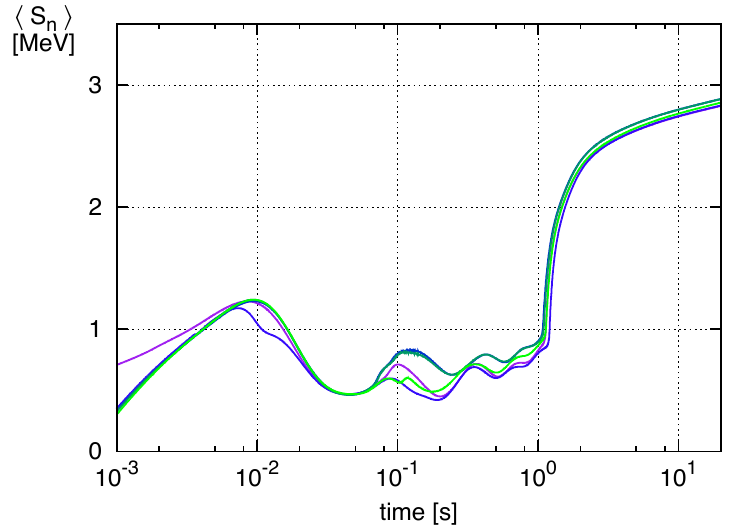}\\ 
    \includegraphics[width=0.48\textwidth]{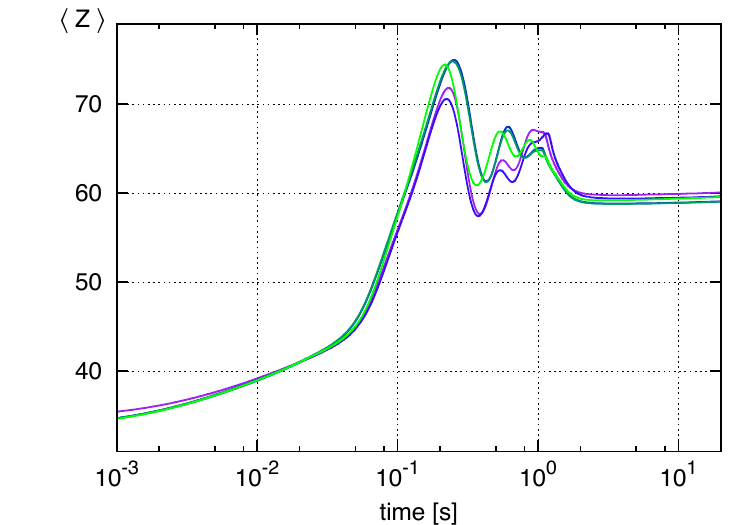}
  \caption{
  Evolution of the neutron density $N_n$ (top), average neutron
  separation energy $\langle S_n \rangle$ (middle) and average proton
  number $\langle Z \rangle$ (bottom) for five trajectories from run 12
  with different initial densities (in the range  
  $10^{12}-10^{13}$~${\rm g}/{\rm cm}^3$).
  }
  \label{fig:nn-sn-z}
\end{center}
\end{figure}
\subsection{Hydrodynamic trajectories}
\label{sec:network}
Our nucleosynthesis calculations are performed with a large reaction
network \citep{winteler12,winteler12b} that is based on the BasNet
network~\citep{thielemann11}. It includes over 5800
isotopes from nucleons up to $Z=111$ between the neutron drip line
and stability.  The reaction rates are from the compilation
of~\cite{rauscher00} for the finite range droplet model
(FRDM)~\citep{moeller95} and the weak interaction rates
(electron/positron captures and $\beta$-decays) are the same as
in~\cite{arcones11b}. In addition, neutron capture and neutron-induced
fission rates of \cite{panov10}  and $\beta$-delayed
fission probabilities as described in Panov et al. (2005) are used.\\
The thermodynamical conditions are taken from hydrodynamical
trajectories of individual SPH particles.  We calculate
nucleosynthetic yields for $30$ representative trajectories in each of
the $21$ ns$^2$ merger simulations, and for $20$ trajectories in the
two nsbh merger simulations. These trajectories are a fair 
representation of the total ejected mass, see Tab.~\ref{tab:runs}.
\\
Once matter has been ejected from the central high-density region, both
temperature and density decrease rapidly while the electron fraction
$Y_e$ can, in principle, start to increase due to $\beta$-decays and
positron captures. 
A small fraction of the trajectories (less than $0.008$, 
see Fig.~\ref{fig:ye-hist}, left panel) suffers shocks during their later evolution
during which their $Y_e$ increases up to $\sim 0.25$, but for
the bulk of trajectories the $Y_e$-values remain essentially 
constant. Some of the higher \ye-material also comes from
the (marginally resolved) neutron star crust material. Although
heavily dominated by values around 0.03, our ejecta electron fractions
show a broader range than those of \cite{goriely11a}, probably due to
our higher resolution (crust material) and their omission of weak
interactions.
\\
We start our nucleosynthesis calculations when the temperature of the
trajectory has dropped below $T_0=10$~GK. The initial composition is
given by nuclear statistical equilibrium and it is further evolved with
the complete network.  Because most of the
trajectories from hydrodynamical simulations terminate after
$t_{\textrm{fin}}\sim10-20$~ms, we extrapolate according to a free 
expansion for the density and an adiabatic expansion law for the
temperature:
\begin{align*}
  \rho(t) = \rho_{\rm fin} \left(\frac{t}{t_{\rm fin}}\right)^{-3},
  \;\;\;
  T(t) = T\left[S_{\rm fin},\rho(t),Y_e(t)\right],
\end{align*}
where $\rho_{\rm fin}$ and $S_{\rm fin}$ are the density and entropy at
$t_{\rm fin}$, while the temperature is calculated at each point of
time from the equation of state \citep{timmes00a}.
\\
Contrary to neutrino-driven winds where the ejecta are dominated
by alpha particles and only few seed nuclei participate in
the r-process, essentially all matter in neutron star mergers 
undergoes r-process. Therefore the r-process energy generation is 
non-negligible and leads to a substantial temperature increase
\citep[see e.g.,][]{freiburghaus99b,metzger10b,goriely11a} that could
affect the late-time evolution \citep{metzger10b}.
We calculate the r-process heating in a post processing step and 
consider its influence on the nucleosynthesis by modifying the temperature.
To account for neutrino energy losses associated with $\beta$-decays,
we introduce a heating efficiency parameter $\epsilon_{th}$
which measures the fraction of nuclear power which is retained in the matter.
\cite{Metzger10a} argue that this fraction must be $\epsilon_{th}\approx0.25-1$. 
As a default, we use $\epsilon_{th}= 0.5$, but below we explore how
(in)sensitive the results are to this choice.
For a given trajectory we take its initial entropy $S_{\rm hyd}(t)$ and
increment it by $\epsilon_{th}S_{\rm nuc}(t)$, following the prescription
of \cite{freiburghaus99b}. 

\subsection{Final abundances}
\label{sec:results}
 
The r-process nucleosynthesis in compact binary mergers is extremely robust,
all 23 cases of Tab.~\ref{tab:runs} deliver essentially identical abundance distributions.
We illustrate the origin of this robustness using the reference case 
of two neutron stars with 1.4~\Msun each (run 12 in
Table~\ref{tab:runs}). Fig.~\ref{fig:standard_results} shows the 
density and temperature evolution for all trajectories (upper panel) 
and their final abundances. Between different trajectories of a single run there is
only a very small spread in the resulting abundances, although the
evolution of temperature is not always the same. The resulting final 
abundances (sum over all ejecta trajectories for each run) are practically the same
for all cases. We show a representative selection of the results from different runs 
in the bottom panel of Fig.~\ref{fig:standard_results}.\\
The robustness of the abundances is mainly due to two factors:
1) the r-process path always reaches the drip line because of the extremely
neutron-rich conditions and
2) there are several fission cycles~\citep[similar to what was observed in
][]{goriely11a}.
These conditions lead to the evolution of the neutron density shown
in Fig.~\ref{fig:nn-sn-z}, upper panel. The huge values of the neutron 
density are due to the (narrow range of) very low \ye-values.\\
In our calculations the low $Y_e$ results in an initial
composition consisting of neutron-rich nuclei ($Z>20$, on the 
neutron drip line) and neutrons. Differences with \cite{goriely11a}
in lighter heavy elements are not particularly relevant because 
these can also be produced in other astrophysical events, and 
the overall yields of the elements with $A < 120$ that we 
obtain are much smaller compared to the heavy robust r-process
elements (similar to \cite{goriely11a}). Note that the contribution 
of the high-$Y_e$ trajectories to the final abundances is negligibly small.
\\
In order to understand the evolution of the r-process path we monitor the
average neutron separation energy (middle panel
Fig.~\ref{fig:nn-sn-z}) defined as
\begin{equation}
  \langle S_n \rangle = \frac{\sum_{Z,A} S_n(Z,A) Y(Z,A)}{\sum_{Z,A}Y(Z,A)},
\end{equation}
with $S_n(Z,A)$ and $Y(Z,A)$ being the neutron separation energy and abundance 
of the nucleus $(Z,A)$. The average separation energy decreases when matter 
moves away from stability and it is, by definition, zero at the neutron drip 
line. The second panel of Fig.~\ref{fig:nn-sn-z} shows that the average 
separation energy is initially below $\approx 1$ MeV, which indicates that 
the r-process path proceeds along the neutron drip line.  
The average proton number increases to $Z=40$ at $t\approx 10^{-2}$s
where the neutron separation energy reaches a maximum. This local 
maximum occurs when the magic number $N=82$ is overcome. Whenever the 
r-process path reaches a neutron magic 
number, it  moves closer to the line of $\beta$-stability (i.e., larger $S_n$ 
values) by increasing $Z$ without changing $N$. After the matter flow passes 
$N=82$ (here around $Z=40$), the $\langle S_n \rangle$ decreases because the path gets again 
further away from the $\beta$-stable region. This continues until the next magic number,  
$N=126$, is encountered (corresponding to the minimum of $\langle S_n \rangle$ between 
0.01s and 0.1s). Around $t= 0.1$s and $Z=60$, $N=126$ is also 
overcome as indicated by the second maximum of $\langle S_n \rangle$. 
After this point the  oscillations in $\langle S_n \rangle$ are due mainly 
to fission as can be seen by the behavior of $\langle Z \rangle$. Note that 
FRDM predicts a magic number at $N=184$, 
consequently the path reaching this point may lead to oscillations in 
$\langle S_n \rangle$.  Therefore, the quantity $\langle Z \rangle$ is 
a nuclear fission indicator that is better suited 
than the average mass number \citep{goriely11a}, since 
it can decrease only through fission reactions, while the average mass number 
can also decrease due to photo-dissociations. The maximum $\langle Z \rangle$
corresponds to the moment when a significant amount of matter has reached the 
region where fission becomes important. The daughter nuclei resulting from 
fission capture neutrons move first towards the drip line and then 
to higher $Z$ where fission acts again. In that way several fission cycles 
occur and lead to oscillations in $\langle Z \rangle$. In Fig.~\ref{fig:nn-sn-z} 
one can distinguish at least three fission cycles. The final increase of 
$\langle S_n \rangle$ is due to the r-process freeze-out at $t \approx$ 1s 
when the neutron-to-seed ratio drops below unity and the matter $\beta$-decays 
to stability. We have used our reference case of two 1.4 \Msun neutron stars, 
run 12, for illustration purposes, but behaviour is very similar 
 in all of the other cases.\\
  \begin{figure*}
  \begin{center}
  \centerline{
  \includegraphics[width=0.52\textwidth]{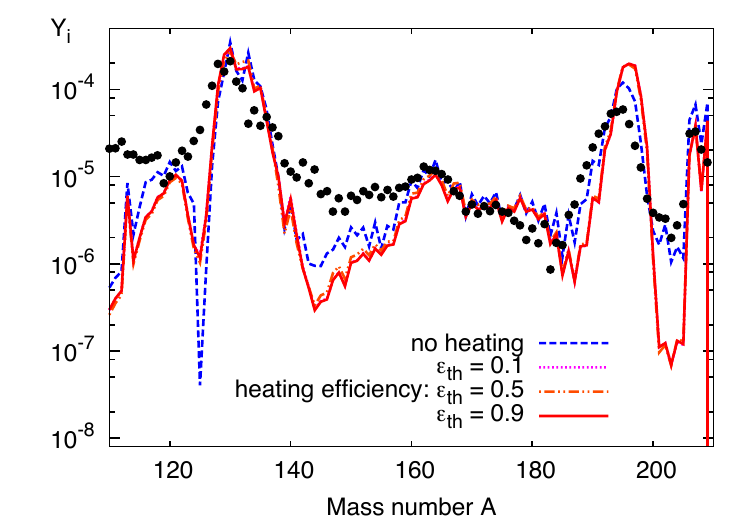}
  \includegraphics[width=0.52\textwidth]{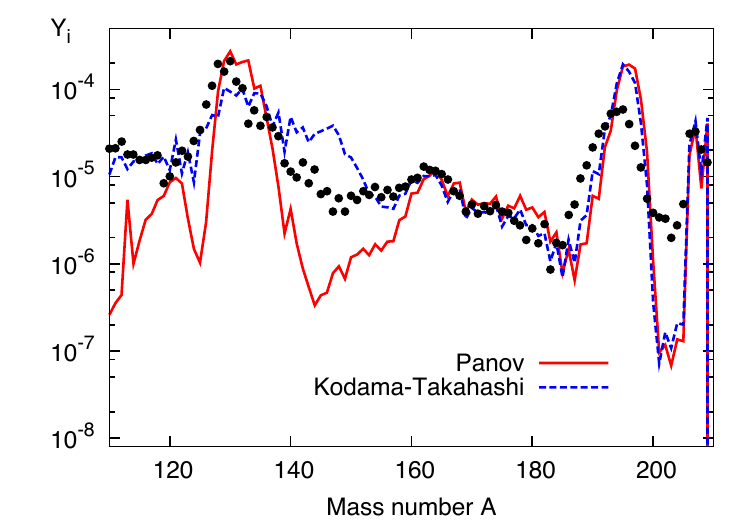} }
  \caption{
    Distribution of final abundances in the vicinity of the second and third
    r-process peaks.
    Left: for the case without heating and for three different values of 
    an heating efficiency $\epsilon_{th}$;
    Right: for two types of fission fragments mass distribution.
    Overplotted black dots represent solar abundances.
  }
  \label{fig:varyparams}
  \end{center}
  \end{figure*}
The major result of this study is that the r-process abundances for compact 
binary mergers do not depend on the astrophysical parameters of the merging
system. They do depend, however, on the not-so-well-known properties of nuclei
close to the neutron drip line. The most critical inputs are the nuclear masses, 
which determine the location of the drip line and the evolution of the path, 
and the fission barriers and yields distribution, which are responsible for
the abundances between $120<A<170$.
We explicitely explore how sensitive our results are to the details of
our nuclear physics input. 
In Fig.~\ref{fig:varyparams}, left panel, we show the effect of including
heating with different efficiencies of the profile of final abundances.
In all cases the second and
third r-process peak are robustly reproduced, but the regions behind
the peaks are less populated if heating is included. In the middle
panel we vary the fraction of nuclear energy that is retained in the
matter, $\epsilon_{th}$. As long as $\epsilon_{th}$ is above some threshold
value ($\sim 1$\%) the resulting abundances are independent of the
exact choice. In the right panel we explore the sensitivity to the 
fission prescription. We once follow  \cite{panov01} assuming only two
fission daughter nuclei and, alternatively, we adopt the fission 
treatment and yield distribution of  \cite{kodama75}. When using the 
latter, the trough after the second peak completely fills up while 
for the Panov prescription the fission products end up entirely in 
the second peak. Note that the used prescription of Panov
uses only two daughter nuclei while he also provides yield distributions in a 
more recent paper \citep{panov08}. Therefore the deviation of our results
from the solar system abundances are sensitive to the 
physics of extremely neutron-rich nuclei.\\
  \begin{figure*}
  \begin{center}
  \includegraphics[width=0.48\textwidth]{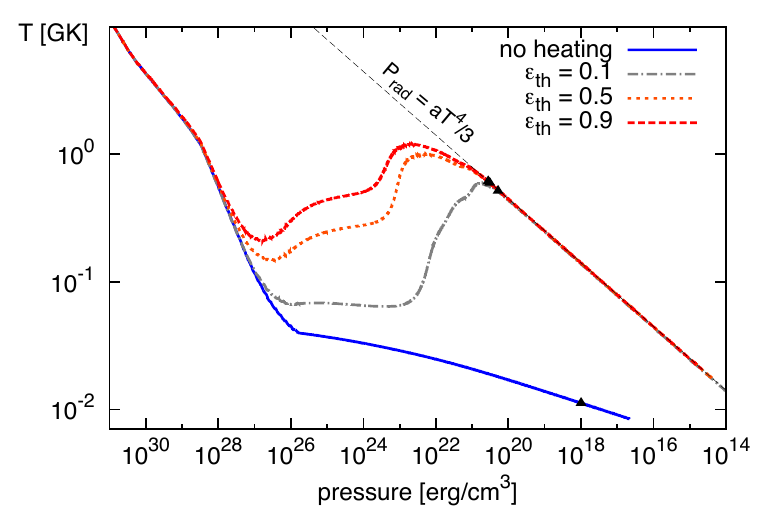}
  \includegraphics[width=0.51\textwidth]{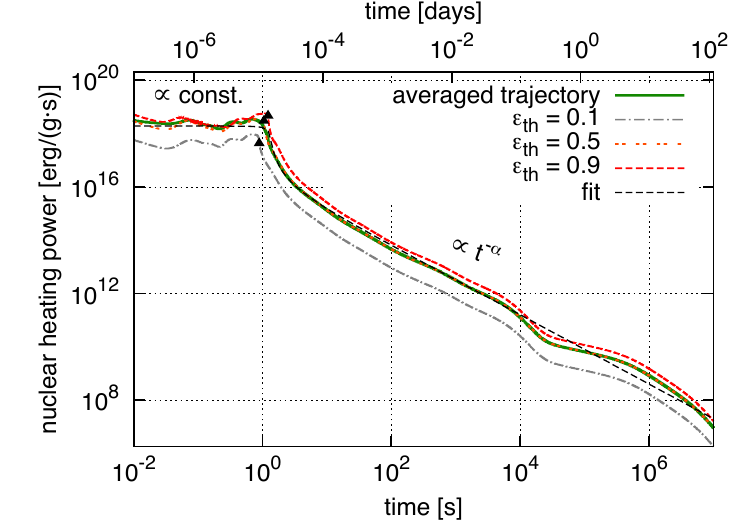}
  \caption{
  Left:
    postprocessed temperature as a function of pressure for the cases
    without nuclear heating (dashed line) and with heating efficiencies
    $\epsilon_{th}=0.1$, $0.5$ and $0.9$.  
  Right:
    nuclear heating power as a function of time, along with a power-law
    fit $\propto t^{-\alpha}$ with index $\alpha=1.3$.
  The small black triangles on each trajectory are neutron freeze-out points 
  (the points in which $Y_n/Y_{\rm seed}=1$).
  }
  \label{fig:heating}
  \end{center}
  \end{figure*}
In Fig.~\ref{fig:heating}, left panel, we show the evolution
of a trajectory in the pressure-temperature plane for different
heating efficiencies, the right panel shows the generated nuclear 
power. The early evolution is very similar for all four cases 
because the internal energy is high and the contribution from the 
generated heat is negligible.
The latter becomes the dominant part of the internal energy after 
about $20$~ms when the temperature and density decrease enough 
for r-process to begin. Similar to ~\cite{freiburghaus99b}
we find that the generated nuclear energy raises the temperature 
by almost an order of magnitude compared to the case without heating.
Matter essentially stays in $(n,\gamma)$-$(\gamma,n)$ equilibrium
until the neutron-to-seed ratio drops below unity (``freeze-out'').
Note that all non-zero heating efficiencies produce trajectories
that lie on the $P=1/3aT^4$-line at freeze-out and therefore follow
a similar behaviour in their decay to stability. 
The reason is that for our extremely neutron-rich conditions
the main contribution to the pressure comes from neutrons and 
photons, while the electron pressure is orders of magnitude smaller. 
Therefore, at the freeze-out point when most of the neutrons are 
captured the system is radiation pressure dominated\footnote{At higher temperatures 
the radiation pressure must necessarily include contribution from 
electron-positron pairs~\citep{witti94,farouqi10}.}.\\
Later the system evolution enters the radiation-dominated expansion which was
studied in~\cite{li98}. In this regime the generated nuclear heat comes mainly 
from $\beta$-decays and can overall be well approximated by a power law, 
while before the freeze-out the generated heat stays approximately constant 
(see Fig.~\ref{fig:heating}). We suggest the following fit which smoothly 
interpolates between a constant value and a power law:
\begin{equation}
  \dot{\epsilon}(t) = \epsilon_0\left(\frac{1}{2}-\frac{1}{\pi}
                  \arctan{\frac{t-t_0}{\sigma}}\right)^\alpha
                  \times\left(\frac{\epsilon_{th}}{0.5}\right)
\end{equation}
with $\epsilon_0=2\times10^{18}$~erg/(g\;s), $t_0=1.3$~s, $\sigma=0.11$~s, and 
$\alpha=1.3$. This formula yields a good fit to all heating histories.
Similarly to the trajectory considered in~\cite{metzger10b}, in our case
starting from $\sim10$~s the nuclear heating power becomes essentially
independent from the initial conditions.
The deviations from a power law that are visible in the right panel of 
Fig.~\ref{fig:heating} at $t\sim10^{-1}$~d and $t\sim20$~d are due to the 
decay of individual elements.\\
%
To explore the dependence of the final abundances
on the electron fraction $Y_e$, we show in Fig.~\ref{fig:vary_Ye}
the resulting abundances for some of the few trajectories 
with higher $Y_e$-values (colour; for comparison the results 
from the dominating low-Ye trajectories are shown in gray). 
For $Y_e$ up to $\sim 0.15$ the results hardly deviate from 
each other, only at and above this value do the abundances 
below $A \sim 110$ start to deviate. Substantial deviations 
for the abundance pattern of the heaviest nuclei with $A>110$ 
only occur for $Y_e > 0.2$, since here the $Y_e$ increase is 
triggered by shocks at low density, the neutron density drops
earlier and the neutron captures freeze out.\\
  \begin{figure}
  \begin{center}
  \centerline{
  \includegraphics[width=0.52\textwidth]{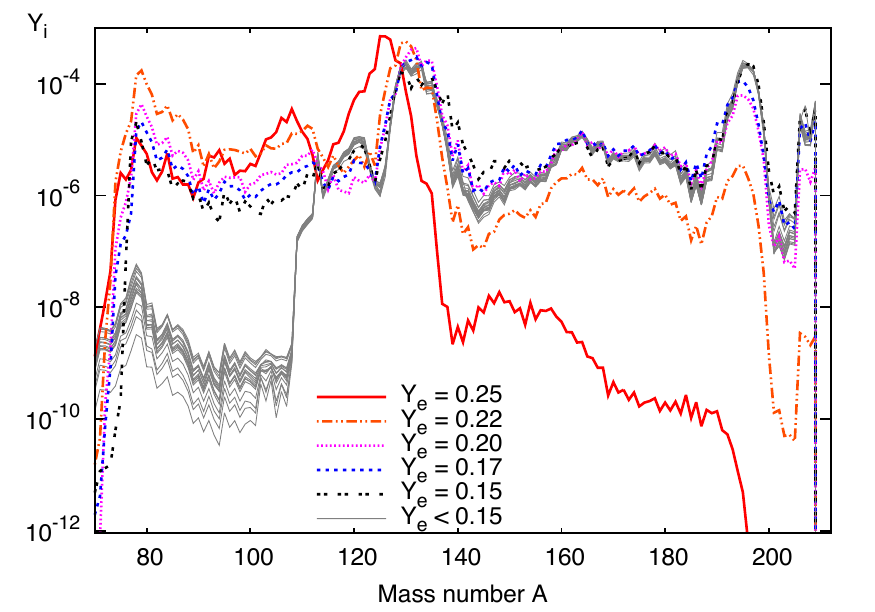} }
  \caption{Dependence of the final abundances on the electron 
           fraction. Shown are the final abundances of some trajectories 
           with high $Y_e$ (in colour), they are compared to the  
           low-$Y_e$ trajectories (gray) that dominate the ejecta. 
           Below $Y_e= 0.1$ two or more fission cycles occur, one or two 
           are realized for $0.1 < Y_e < 0.22$ and none for
           $Y_e > 0.22$. Since high-$Y_e$ trajectories only represent a 
           very small mass fraction, their contribution to the overall 
           abundances is minuscule.
  }
  \label{fig:vary_Ye}
  \end{center}
  \end{figure}
\section{Summary and discussion}
\label{sec:discussion}

In this study we have re-examined the question of heavy element nucleosynthesis
in compact binary mergers. Our study adds several new aspects in comparison
with earlier work. First, we systematically cover the plausible
ns$^2$ parameter space with masses from 1.0 to 2.0 \Msun in 21 simulations. 
Despite the long history of ns$^2$ simulations we are not aware of any study
that has systematically explored such a wide range of ns masses. We complement
these cases with mergers of a 1.4 \Msun neutron star with black holes of 5 and 
10 \msun. Second, since we use a nuclear equation of state with neutron stars 
in initial cold $\beta$-equilibrium and include \ye-changing weak interactions, 
we overcome the \ye-ambiguity that has plagued all previous studies.
We subsequently find the final abundances by following the nucleosynthesis 
along a large set of
hydrodynamic trajectories with a state-of-the-art nuclear reaction network.\\
  \begin{figure}
  \centerline{
  \includegraphics[width=0.52\textwidth,angle=0]{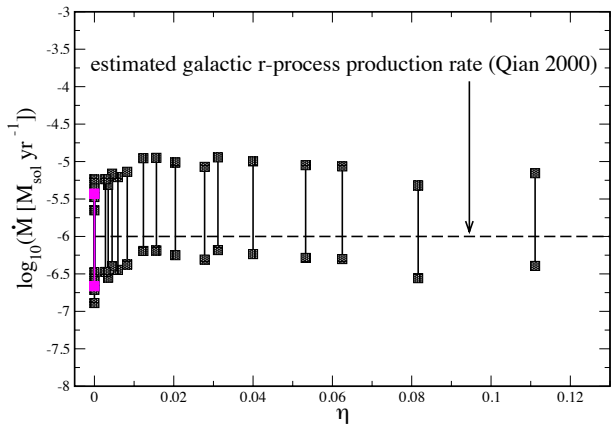}}
  \caption{Relevance of the ejected mass for the r-process enrichment of the
           galaxy. The mass asymmetry parameter $\eta$, see Eq.~(\ref{eq:eta}),
           is used here to label the different simulations. 
           Adopting the 95\% confidence intervall for the galactic 
           double neutron star merger rate from \citep{kalogera04c}, 
           we calculate
           which galactic enrichment rates would result for the ejected matter of our
           simulations (lines with squares at ends; our reference case of two 1.4~\Msun
           neutron stars is shown in magenta). For comparison we also plot the estimated
           galactic enrichment rate \citep{qian00}. All the ejected masses are consistent
           with being a major contributor to galactic r-process, even if the total amount
           per event should be decreased by a factor of a few (say, due to equation of state
           or GR effects).}
  \label{fig:ejmass}
  \end{figure}
Consistent with earlier studies, we find a very robust r-process and final 
abundances that are in good agreement with the heavy solar system abundance pattern.
The major new result is that the final pattern is extremely robust accross
the whole parameter space: all 21 ns$^2$-merger and the two nsbh-merger cases 
yield practically identical nucleosynthesis outcomes. The major reason for this
unique abundance pattern is the extreme neutron richness of the ejecta, 
$\langle Y_e\rangle\approx 0.04$. Consequently, in each case the r-process path meanders along
the neutron drip line and matter undergoes several fission cycles, so that the abundances
are determined entirely by nuclear rather than by astrophysical properties.
As a corollary, the poorly known nuclear properties near the neutron dripline do have
an impact on the resulting abundance pattern. We find some dependence on the used mass formula
and on the distribution of the nuclear fragments after fissioning. Nevertheless, the 
second and third r-process peaks are robustly reproduced, without any ``tuning'' of the
nuclear physics input the overall agreement with the solar system r-process pattern is good.
R-process matter lighter than $\sim A\approx 120$, however, is substantially 
under-produced with respect to the solar system pattern. The few high-$Y_e$ 
trajectories produce different abundance patterns, see Fig.~\ref{fig:vary_Ye}, 
but there is too little mass in this material to leave a noticable impact in the
resulting average distribution.\\
One may wonder whether captures of neutrinos from the central remnant, 
which are not included in our simulations, could change the \Ye of the ejecta
noticeably. For two reasons, we do not think that this is the case. First, 
the ejecta are launched practically at first 
contact, see Fig.~\ref{fig:escaper_region}, with $\langle v \rangle > 0.1 c$, 
\citep{piran12a}. The neutrino emission, however, only becomes relevant after 
a disk has formed which takes $\sim 7$ more milliseconds (e.g. Fig. 7 in \cite{rosswog12a}) 
at which time the solid angle under which the escaping matter is seen has 
decreased already by a large factor. Second, the ejecta are concentrated in 
the orbital plane while the neutrinos are emitted preferentially perpendicular 
to it, see Fig. 12 in \cite{rosswog03a} and Fig. 9 in \cite{dessart09}. We 
therefore expect \ye-changes due to neutrino captures to be 
negligible\footnote{This may be different for likely emerging neutrino-driven 
winds that might also contribute to the nucleosynthesis of compact binary merger 
events.}.\\
The dynamic ejecta masses are large enough to make a substantial contribution 
to the chemical enrichment of the galaxy with r-process material. 
Adopting the observation-based double neutron star merger rates of \cite{kalogera04c}, 
$R_{\rm DNS}= 83.0 ^{+209.1}_{-66.1}$ yr$^{-1}$, we calculate by which average rate the galaxy 
would be r-process enriched from neutron star mergers. Although 
the overall ejecta masses vary by more than a factor of five between different cases, all of them
are perfectly consistent with the estimated galactic enrichment rate \citep{qian00}, see 
Fig.~\ref{fig:ejmass}. The reference
case of $2 \times 1.4$ \Msun neutron stars (shown in magenta), nicely brackets this
value. This also implies that the relevance of compact binary mergers for r-process remains
unaffected even if the typical ejecta amount per event should be decreased by a factor of a few, 
say, due to equation of state or GR effects.\\
Our findings of ejecta masses in the required range, extremely robust abundance patterns for the heaviest
r-process elements, but underproduction of the lighter r-process elements, make compact binary mergers
the major candidate for the observed, unique, heavy r-process component \citep{sneden08}. At the same time
this means, that at least one, but possibly more additional r-pocess sites must contribute to the 
cosmic chemical evolution of lighter r-process elements.\\ 
An important question is how early the r-process yields from compact binary mergers
would be available for cosmic enrichment. The inspiral time from the birth of a compact binary system
to a coalescence is very sensitive to both the initial sparation and the orbital eccentricity \citep{peters63,peters64}.
The initial separation is set by binary evolution processes which are beyond the scope of this 
paper, but it is worth re-iterating that a binary system that survives the two SN explosions that are
required to form the system in the first place will in most cases possess a large orbital eccentricity. 
As illustrated in Fig.~\ref{fig:lifetimes} short lifetimes can be achieved with plausible orbital 
parameters. For example, an initial semi-major axis of 1 R$_\odot$ and an eccentricity of 0.9 
lead to a merger only 1 Myr after binary formation (for comparison we note that the projected 
semi-major axis of the double pulsar PSR J0737-3039 A+B \citep{burgay03,kramer08} is 0.6 R$_\odot$). 
Some binary population synthesis studies \citep{dewi03,belczynski06} find indeed short-lived channels
that coalesce essentially {\em at} birth, so an early enrichment from compact mergers is at least plausible.
Earlier studies \citep{argast04} disfavoured neutron star mergers as dominant 
galactic r-process source. Based on our hydrodynamic plus nucleosynthesis studies, 
however, we have shown that compact binary mergers are excellent production sites 
for the unique, heavy r-process component. We leave the question of whether or to which 
extent they are consistent with cosmic chemical evolution for future studies.\\

  \begin{figure}
  \centerline{
  \includegraphics[width=0.52\textwidth]{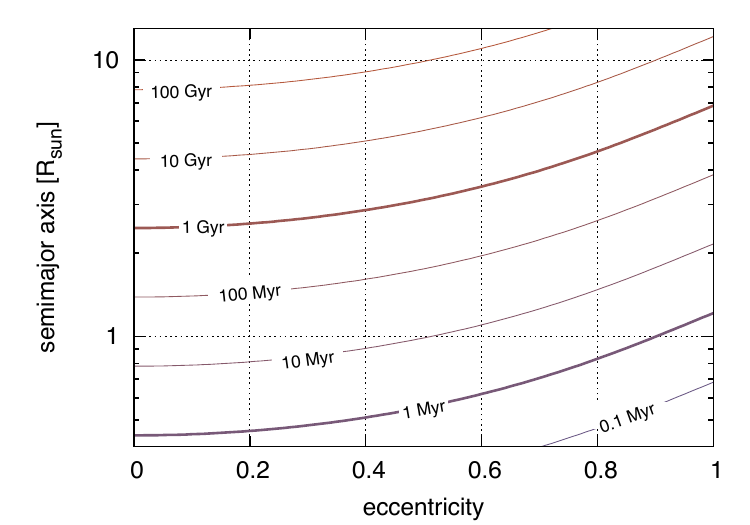}}
  \caption{Contour plot of the estimated lifetimes of a 1.4-1.4~\Msun binary
           for different orbital eccentricities and semimajor axes~\citep{peters64}.}
  \label{fig:lifetimes}
  \end{figure}

\section*{Acknowledgments}

This work was supported by DFG grant RO-3399, AOBJ-584282.
S.R. thanks Enrico Ramirez-Ruiz/UC Santa Cruz and Andrew MacFadyen/NYU
for their hospitality and for many stimulating conversations. We 
also thank Gabriel Martinez-Pinedo for useful discussions on r-process
nucleosynthesis. Our particular thanks goes to K.-L. Kratz and F.-K. 
Thielemann for their continued support and interest in our work. 
Especially FKT has educated us on this exciting topic and we are
grateful for his insightful comments on an earlier version of this manuscript.
We acknowledge the use of the visualization software SPLASH
developed by Daniel \cite{price07d}. The simulations of this
paper were performed on the facilities of the
H\"ochstleistungsrechenzentrum Nord (HLRN). A. A. is supported by the
Helmholtz-University Young Investigator Grant No. VH-NG-825. \\

\bibliography{astro_SKR}
\bibliographystyle{mn2e.bst}
\bsp

\label{lastpage}

\end{document}